\documentclass{emulateapj}

\usepackage{amssymb}
\usepackage{amsmath}

\def\apj{\rm{ApJ}}                 
\def\apjl{\rm{ApJ}}                
\def\apjs{\rm{ApJS}}                        
\def\mnras{\rm{MNRAS}}   

\usepackage{epsfig}
\shorttitle{Star Formation From SPLASH}
\shortauthors{Steinhardt et al.}

\begin{document}


\title{Star Formation at $4 < z < 6$ From the Spitzer Large Area Survey with Hyper-Suprime-Cam (SPLASH)}


\author{Charles L. Steinhardt \altaffilmark{1,2,3}, Josh S. Speagle \altaffilmark{3,4}, Peter Capak \altaffilmark{1,2},  John D. Silverman \altaffilmark{3}, Marcella Carollo \altaffilmark{5}, James Dunlop \altaffilmark{6}, Yasuhiro Hashimoto \altaffilmark{7}, Bau-Ching Hsieh \altaffilmark{8}, Olivier Ilbert \altaffilmark{9}, Olivier Le Fevre\altaffilmark{9},  Emeric Le Floc'h \altaffilmark{10}, Nicholas Lee \altaffilmark{11}, Lihwai Lin \altaffilmark{8}, Yen-Ting Lin \altaffilmark{8}, Dan Masters \altaffilmark{1,2}, Henry J. McCracken \altaffilmark{12}, Tohru Nagao \altaffilmark{13}, Andreea Petric \altaffilmark{1,14}, Mara Salvato \altaffilmark{15}, Dave Sanders \altaffilmark{11}, Nick Scoville \altaffilmark{1}, Kartik Sheth \altaffilmark{2,16}, Michael A. Strauss \altaffilmark{17}, Yoshiaki Taniguchi \altaffilmark{18}}

\altaffiltext{1}{Caltech, 1200 E. California Blvd., Pasadena, CA 91125, USA}
\altaffiltext{2}{IPAC, 1200 E. California Blvd., Pasadena, CA 91125, USA}
\altaffiltext{3}{Kavli IPMU, University of Tokyo, Kashiwanoha 5-1-5, Kashiwa-shi, Chiba 277-8583, Japan}
\altaffiltext{4}{Harvard University Department of Astronomy, 60 Garden St., MS 46, Cambridge, MA 02138, USA}
\altaffiltext{5}{ETH Zurich, Wolfgang-Pauli-Strasse 27, 8093 Zurich, Switzerland}
\altaffiltext{6}{Institute for Astronomy, University of Edinburgh, Royal Observatory, Blackford Hill, Edinburgh EH9 3HJ, UK}
\altaffiltext{7}{National Taiwan Normal University, No. 88, Sec. 4, Tingzhou Rd., Taipei 11677, Taiwan R.O.C.} 
\altaffiltext{8}{ASIAA Sinica, AS/NTU. No. 1, Sec. 4, Roosevelt Rd, Taipei 10617, Taiwan, R.O.C.}
\altaffiltext{9}{Laboratoire d'Astrophysique de Marseille, 38 rue Frederic Joliot Curie, 13388 Marseille, France}
\altaffiltext{10}{Service d'Astrophysique, CEA-Saclay, Orme des Merisiers, Bat. 709, 91191 Gif-sur-Yvette, France}
\altaffiltext{11}{Institute for Astronomy, 2680 Woodlawn Drive, Honolulu, HI 96822, USA}
\altaffiltext{12}{Institut d'Astrophysique de Paris, 98 bis boulevard Arago, 75014 Paris, France}
\altaffiltext{13}{Kyoto University, Kitashirakawa-Oiwake-cho, Sakyo-ku, Kyoto 606-8502, Japan}
\altaffiltext{14}{Gemini Observatory, 670 N. A'ohoku Place, Hilo, Hawaii 96720, USA}
\altaffiltext{15}{Max Planck Institute for Extraterrestrial Physics (MPE), Giessenbachstr. 1, D-85748 Garching. Germany}
\altaffiltext{16}{North America ALMA Science Center, National Radio Astronomy Observatory, 520 Edgemont Road, Charlottesville, VA 22903, USA}
\altaffiltext{17}{Princeton University, Dept. of Astrophysical Sciences, Princeton, NJ 08544, USA}
\altaffiltext{18}{Ehime University, 10-13 Dogo-himata, Matsuyama, Ehime 790-8577, Japan}


\begin{abstract}

Using the first 50\% of data collected for the Spitzer Large Area Survey with Hyper-Suprime-Cam (SPLASH) observations on the 1.8 deg$^2$ Cosmological Evolution Survey (COSMOS) we estimate the masses and star formation rates of 3398 $M_*>10^{10}M_\odot $ star-forming galaxies at $4 < z < 6$ with a substantial population up to $M_* \gtrsim 10^{11.5} M_\odot$.  We find that the strong correlation between stellar mass and star formation rate seen at lower redshift (the ``main sequence'' of star-forming galaxies) extends to $z\sim6$.  The observed relation and scatter is consistent with a continued increase in star formation rate at fixed mass in line with extrapolations from lower-redshift observations.  It is difficult to explain this continued correlation, especially for the most massive systems, unless the most massive galaxies are forming stars near their Eddington-limited rate from their first collapse.  Furthermore, we find no evidence for moderate quenching at higher masses, indicating quenching either has not occurred prior to $z \sim 6$ or else occurs rapidly, so that few galaxies are visible in transition between star-forming and quenched.

\end{abstract}



\section{Introduction}

Studies of star-forming galaxies at a wide range of cosmic epochs have revealed a strong correlation at fixed redshift between star formation rate (SFR) and stellar mass ($M_*$), of the form
\begin{equation}
\log SFR (M_\odot/\textrm{yr}) = \alpha \times (\log M_*/M_\odot - 10.5) + \beta,
\label{eq:sfr}
\end{equation}
with $\alpha$ and $\beta$ likely time-dependent.  This relationship has been shown to hold over 4\,--\,5 orders of magnitude in mass \citep{santini+09} and from $z = 0$ to $z \sim 6$  \citep{noeske+07,daddi+07,karim+11,kashino+13,Speagle2014}.  It is a tight correlation, with only $\sim .25$\,--\,$.35$\,dex of scatter at any redshift \citep{daddi+07,Whitaker2012}.  

This paper extends the star-forming main sequence (SFMS) to massive galaxies at higher redshifts than previous studies by using the first data from the Spitzer Large Area Survey with Hyper-Suprime-Cam (SPLASH).  This survey is obtaining 2475h ($>$6h/pointing) of Spitzer IRAC 3.6$\mu$m and 4.5$\mu$m observations on the two Hyper-Suprime-Cam \citep{Takada2010} ultra-deep fields COSMOS \citep{scoville+07,Sanders2007} and SXDS \citep{SXDS}. Here we use the first 50\% of the data on the COSMOS field ($\sim4$h/pointing) along with previously published 0.15-2.5$\mu$m data \citep{ilbert+13} to probe star forming galaxies with masses $>10^{10}M_\odot$ to z$\sim6$.

SPLASH is is a photometric survey that improves upon existing studies at $4 < z < 6$.  For these galaxies, the Lyman break allows quality photometric redshift determination, while SPLASH is complete and deep enough to greatly reduce biases that occur in studies specifically selecting for Lyman break galaxies \citep{lee+12,Speagle2014}. 

In \S~\ref{sec:dataset}, we describe the SPLASH catalog, as well as the spectral energy distribution (SED) template fitting we use to produce photometric redshifts (photo-z's), stellar masses, and SFRs.  \S~\ref{sec:mainseq} describes the SPLASH view of the SFMS, including a lack of observed quenching at the high-mass end.  Possible explanations for this lack of a turnoff are discussed in \S~\ref{sec:discussion}, and in more detail in \citet{Steinhardt2014}.

Throughout this work, we assume a standard $(H_0,\Omega_m,\Omega_\Lambda)$ = $(70,0.3,0.7)$ cosmology, AB magnitudes, and a Chabrier \citep{Chabrier2003} IMF (integrated from $0.1 M_\odot$ to $100 M_\odot$).  

\section{The SPLASH Dataset}
\label{sec:dataset}

This work uses a revised version of the Subaru i$^+$ band selected Cosmological Evolution Survey (COSMOS) catalog from \citet{Capak2007} to provide $0.15-2.5\mu$m photometry, with $10^6$ spurious sources removed \citep{salvato+09}, updated with intermediate band data, photo-z's, and physical parameters as described in \citet{ilbert+10}.  The catalog was further augmented with significantly (1 magnitude) deeper z$^+$ band data taken with an updated Subaru-Suprime-Cam, Ultra-Vista \citep{McCracken2012} imaging in Y, J, H, and K bands, and importantly the SPLASH IRAC $3.6\mu$m and $4.5\mu$m data.  A full description of SPLASH is provided in Capak et al. (in preparation).  

In this paper we use the first 50\% of the SPLASH IRAC data, which covers a 1.2 degree diameter circle centered on the COSMOS field to 5$\sigma$ $\sim 25.3$ mag$_{\textrm{AB}}$ at both wavelengths. To reduce AGN contamination, X-ray detected sources \citep{Brusa2010,Civano2012} were removed from the sample. To overcome source confusion (blending) we extracted the photometry using the $i^+$ band catalog as input for the IRACLEAN procedure described in \citet{Hsieh2012}.  The IRACLEAN photometry was compared with a T-FIT \citep{Laidler2007} catalog in the CANDELS \citep{Grogin2011,Koekemoer2011} area using an HST-WFC3 H band image as a prior and an EMPHOT \citep{Conseil2011} catalog of the whole COSMOS field using the Ultra-Vista K band image as a prior.  No systematic bias as a function of flux or position was found due to the photometric method.

The primary selection effect in our sample is the i-band catalog, which is used as a prior for the Spitzer photometry and as the basis for the photometric catalog.  The use of this catalog limits us to $z \lesssim 6$, when the Lyman break leaves the i-band.  The depth limits us to unobscured star formation rates $>5-10 M_\odot$/yr at $z\sim4-6$ (typical excitations are $A_V \sim 0.4$ mag for objects in our sample; cf. \citet{Bouwens2012}).  The typical depth of i-band selection is 26.5 mag \citep{Capak2007,Ilbert2009}.  The second limit is the depth of the IRAC data, which introduces an age-dependent mass cutoff.  For zero-age galaxies with a flat spectral energy distribution the IRAC data will limit the lowest measurable masses. For older galaxies, however, the depth of the Subaru $i^+$ image will limit the lowest measurable masses since they would not be in the i-band catalog.  For stellar populations with approximately the age of the universe, our joint mass cutoff is noted in Figure 2.  A third, more subtle effect is the method used to extract the IRAC photometry.  If the photometry is not properly deblended, fainter galaxies will have systematically increased fluxes due to photometric crowding.   The IRACLEAN algorithm corrects for this at least as well as other methods (TFIT, EMPHOT), but we note that this may be affecting a small number of objects crowded by a source not in the i-band catalog.  

\subsection{Photometric Redshifts}
\label{subsec:photoz}

We determine photo-z's using Le\_PHARE \citep{Arnouts2011} with the methodology described in \citet{ilbert+13}. Stellar masses were estimated with \citet{Bruzual2003} (BC03) models including strong emission lines with a mix of exponentially-declining ($\tau/\textrm{100 Myr}=[0.1,0.3,1,3,10,30]$) and delayed-$\tau$ ($\Delta t_{\textrm{peak}}/\textrm{Gyr}=[1,3]$) star formation histories (SFHs) with solar and half-solar metallicities at a range of ages spanning 0.05-13.5 Gyr to account for the effects of low metallicities and increasing SFHs on estimated physical parameters. Dust attenuation is modelled using the starburst \citet{calzetti+00} curve and the $\lambda^{-0.9}$ curve from \citet{Arnouts2013} with $E(B-V)$ up to 0.7 mag.

\begin{figure}[!ht]
\plotone{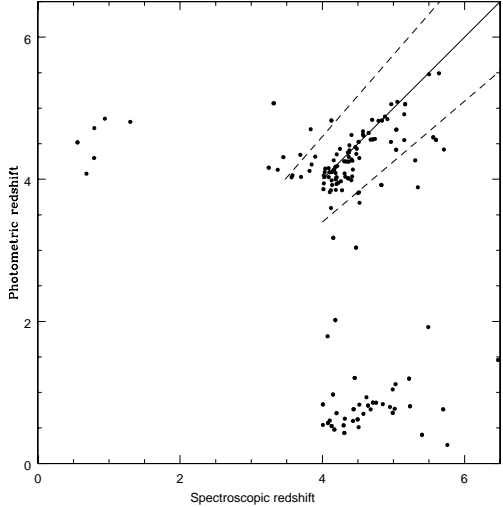}
\caption{A comparison of the best-fit photometric redshift and DEIMOS spectroscopic redshift for objects in SPLASH at $z_{phot} > 4$ or $z_{spec} > 4$.  Due to confusion between the Balmer break and Lyman break, many objects with catastrophic errors are mistakenly excluded from the sample (scattered to $z \sim 0.5$), but very few objects are scattered up, even though there are an order of magnitude more spectroscopic galaxies at low redshift.}
\label{fig:DEIMOScomp}
\end{figure}

Although the presence of a Lyman break should result in precise photo-z's, the 4000\AA\  break and Lyman break can be confused resulting in catastrophic redshift failures.  To check the quality of the photo-z's we cross-reference them with spectroscopically confirmed sources (spec-z's), primarily from DEIMOS observations selected to be representative of $z>4$ galaxies \citep{Capak2010,Mallery2012}.  Out of 139 galaxies with spec-$z > 4$ and robust IRAC fluxes, 87 (63\%) have successful photo-z determinations (Fig. \ref{fig:DEIMOScomp}) to within 15\%, with most of the rest mistakenly fit with at much lower redshift.  Howvever, there are very few false positives in the SPLASH catalog.  6\% of objects with photo-$z > 4$ and measured spec-$z$s lie at spec-$z < 2$ (indicating confusion between the two breaks), and after correcting for the significantly larger fraction of objects with measured spectroscopic redshifts at $z<2$ (10\% at $z<2$ vs 4\% at z$>4$ with ch1$<25.3$), this implies a contamination rate of 2\%.  So, the current photo-z analysis is preferentially scattering objects down from $z \sim 4-6$ to $z \sim 0.5$,  (Fig. \ref{fig:DEIMOScomp}) with approximately 40\% of SPLASH star-forming galaxies erroneously fit as low-redshift and removed from the sample.  Tests using the two-point correlation and cross-correlation functions confirm these fractions (Coupon et al., private communication).

We note that typical SFRs are far lower at $z \sim 0.5$ than $z > 4$, so up-scattered objects fall well off the SFMS at $z>4$ where they will not affect our further analysis.

\subsection{Stellar Masses and Star Formation Rates}

Once the best-fit photo-z's and extinctions are determined, Le\_PHARE fixes both and fits for a stellar mass and SFR based upon the BC03 templates with emission lines added at the best-fit redshift \citep{Ilbert2009}.  We compared these masses to those with several different SFHs, dust extinction laws, and emission line prescriptions similar to \citet{Arnouts2013} and found the results varied by less than 0.2 dex.  Furthermore, we found consistent stellar mass estimates for objects at $z>4$ when comparing these results with \citet{ilbert+13}, with the exception of heavily blended objects where our improved IRAC photometry had an effect.  We note that although stellar mass estimates come from SED fits at all redshifts, at lower redshifts it is possible to measure SFRs independently.  Although beyond the scope of this Letter, this is discussed extensively in \citet{Speagle2014}.  

  At $z\sim0-3$, several studies have investigated SED-based SFRs and found there is generally good agreement between the SED and other SFR indicators  \citep{salim+07,Arnouts2013,Carollo2013}.  To verify this at the higher-redshifts in SPLASH we tested full SED fitting-based SFR indicators against UV- \citep{Meurer+99} and FIR- \citep{Lee2013} based SFRs.  
 
It was possible to derive reasonable UV SFRs using the IR excess - UV slope (IRX-$\beta$) relation \citep{Meurer+99} for objects with detections in at least three bands in the rest-frame 1600-3000 \AA~ range.  Galaxies with either negative $A_{1600}$ or unconstrained $\beta$ were removed, leaving a sample of 404 galaxies.  There is good agreement in the mean values of these UV and SED-based SFRs over more than two orders of magnitude in SFR, but with a large scatter of 0.8 dex.  This scatter is consistent with expectations at these high redshifts \citep{Bouwens2013,Finkelstein2012} due to both the intrinsic properties of these galaxies and measurement error.  

For galaxies with very high SFRs ($\gtrsim500 M_\odot/\textrm{yr}$) we use Herschel fluxes (Lee et al., in prep.) to estimate FIR-based SFRs and find generally good agreement.  The scatter between the SED and FIR based SFRs is also large but consistent with other studies (e.g., \citet{Reddy2008}).  We conclude that our SED based SFRs are reasonable, but may be under-estimating the actual scatter in the SFMS.  One additional concern is that we only use one dust extinction law, but in reality there are likely many, and we cannot disambiguate at these redshifts.  

\subsection{Sample Selection}

There are 2,002,871 sources in the SPLASH catalog, of which 1,319,197 are classified as stars or galaxies with well-constrained redshifts, masses, and SFRs.  Of these, 7583 have a best-fit photo-z $> 4$ (\S~\ref{subsec:photoz}).  We removed any objects where quasar or stellar SEDs had better fits than the galaxy templates.  Assuming conservatively that the stellar population has the same age as the universe, SPLASH is mass complete at $\log M_*/M_\odot > 10$ at these redshifts, a limit that will be improved by over 1 mag with upcoming Hyper-Suprime-Cam (HSC) data in the visible.

To obtain a clean sample we chose a quality cut based on the SED fitting uncertainty, $\chi^2/\textrm{DOF} < 5$, rather than a single band signal-to-noise (S/N).  In practice, this is primarily a selection for having many observed bands with S/N $>3$.  This leaves a sample of 3398 objects between redshifts 4 and 6 (2541/857 above/below $z = 4.8$, dividing the range into two bins of equal cosmic time).  Limiting the sample to the mass-complete regime leaves 2152 objects (1513/639).  We will use this smaller, mass-complete (sub)sample when deriving relationships between physical quantities to avoid biases.  We include in our sample approximately 50 objects with no IRAC detection but otherwise well-constrained SED fits that pass the remainder of our cuts to avoid introducing a bias against particular SED shapes. Removing these objects has a negligible effect on the conclusions drawn in \S~\ref{sec:mainseq}.

Objects selected by the $\chi^2$ cut are preferentially brighter in the optical, and may introduce a bias in the sample distribution.  To test for this we did a Kolmogorov-Smirnov comparison between the inferred SFR and stellar mass of our $\chi^2/\textrm{DOF} < 5$ sample and the original sample.  The two are consistent with being drawn from the same distribution.  

As discussed in \S~\ref{subsec:photoz}, approximately 80 objects ($\sim 2\%$) of our sample are expected to be low-redshift galaxies fit with $z > 4$ due to confusion between the 4000\AA~ and Lyman breaks.  At $z > 4$, it is expected that all galaxies are star-forming \citep{Brammer2011,Behroozi2013}, and $z < 2$ galaxies of any type scattered up to $z>4$ will be assigned much lower SFR or even be selected as quiescent due to the shape of the typical galaxy SED.  73/3398 galaxies have SFRs below 10 $M_\odot$/yr (70 with $z < 4.8$), while 46/3398 are found to be quiescent according to the \citet{ilbert+13} criteria based on dust-corrected NUV$rJ$ criteria.  Because these two selections have little overlap and comprise a negligible fraction of the catalog, we choose to use all 3398 objects in our final sample.

\section{The Star-Forming Main Sequence}
\label{sec:mainseq}

\begin{figure}[!ht]
\plotone{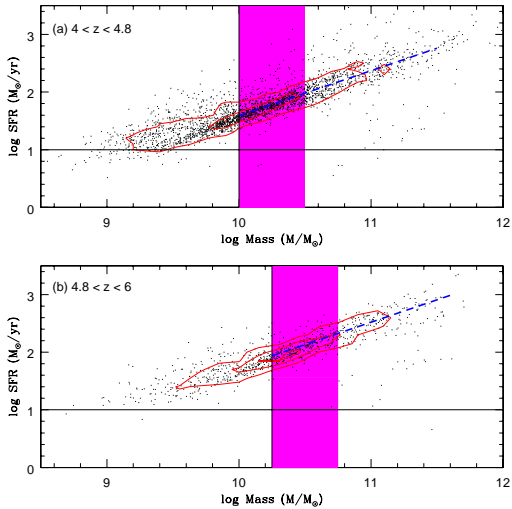}
\caption{The ``main sequence'' for star-forming galaxies at photometric (a) $4 < z < 4.8$ and (b) $4.8 < z < 6$.  A best-fit linear relationship is indicated by the blue dashed line in each panel.  Mass and SFR completeness are indicated by the solid black lines.  The magenta shaded region reflects an estimate of the increased sensitivity of SPLASH due to the addition of IRAC channels to existing multiwavelength data.  Contours are drawn are equal intervals in number density.}
\label{fig:SFfig}
\end{figure}

The main result of this paper, the SFMS from SPLASH at $4 < z < 4.8$ and $4.8 < z < 6$ (bins of equal time interval), are shown in Fig. \ref{fig:SFfig}.  
In both redshift ranges, there is a strong correlation between stellar mass and SFR, qualitatively similar to that seen in approximately two dozen previous studies \citep{Speagle2014} but with a different locus than at lower redshift.  Almost all objects lie near this SFMS, with a small number of clear outliers at low SFR.  Because of the limits of broad-band SED fitting, it is unclear without followup observations whether these outliers are candidates for quenched galaxies, catastrophic errors in photo-z's, and/or SED fitting, or have some alternative explanation.

The main sequence in each panel is divided by stellar mass into bins of width $0.1$ dex, with the median for each bin shown in Fig. \ref{fig:SFturnoff}.  For ease of comparison with other studies, medians above the mass completeness threshold that encompass $> 25$ objects are then fit with a power law (Eq. \ref{eq:sfr}),
with best-fit $\alpha$ and $\beta$ indicated in Table \ref{table:fits}. Using both fits, individual objects exhibit 1$\sigma$ scatter of $\sim 0.24$ dex about the main sequence, and agree well with Lyman-break selected objects from \citet{lee+12} at similar/slightly lower redshifts and lower stellar masses.  
\begin{table*}
\caption{Best-fit Parameters for the Main Sequence at $4<z<6$}
\centering
\begin{tabular}{l|c|c|c|c}
\label{table:fits}
Source & $\alpha(4-4.8)$ & $\beta(4-4.8)$ & $\alpha(4.8-6)$ & $\beta(4.8-6)$ \\
\hline
Measured &  $0.78 \pm 0.02$ & $1.976 \pm 0.005$ & $0.78 \pm 0.02$ & $2.110 \pm 0.003$ \\
Lee+12 &  $0.79$ & $1.955$ & $0.73$ & $1.975$ \\ 
Whitaker+12(z) & 0.13 & 1.72 & 0.02 & 1.17 \\
Speagle+14(t) & $0.79 \pm 0.03$ & $2.11 \pm 0.04$ & $0.80 \pm 0.03$ & $2.16 \pm 0.04$ \\
\end{tabular}
\\
{\scriptsize Best fit slope $\alpha$ and SFR at $\log M/M_\odot = 10.5$ $\beta$ for the star-forming main sequence in SPLASH.  The expected fits according to the redshift and time dependence given by \citet{Whitaker2012} and \citet{Speagle2014}, respectively, are included for comparison, as well as the best fit values for the Lyman-break selected samples of \citet{lee+12} at $z \sim 3.9$ and $5.0$ corrected for extinction via \citet{bouwens+12}.}
\end{table*}

Previous studies have disagreed on whether $\alpha$ is increasing (e.g., \citet{santini+09}), constant (e.g., \citet{karim+11}), or decreasing (e.g., \citet{Whitaker2012}) with redshift.  We find a high-redshift slope of $0.78 \pm 0.02$ (Table \ref{table:fits}), in good agreement with \citet{lee+12} and the increasing-$\alpha$ prediction from the \citet{Speagle2014} literature compilation.  The uncertainties given are statistical, but this good agreement may imply that the unknown systematic uncertainties are smaller than might otherwise have been expected.

\subsection{High-Mass Galaxies and Quenching}  

As star formation is quenched, individual galaxies should turn off the SFMS, either lying at lower SFR or becoming quiescent and disappearing entirely due to selection.  It has been previously suggested that high-mass quenching is observed as a sub-linear SFR-$M_*$ relationship at high masses, but this may be due to selection effects \citep{noeske+07,karim+11,Whitaker2012}.  

\begin{figure}[!ht]
\plotone{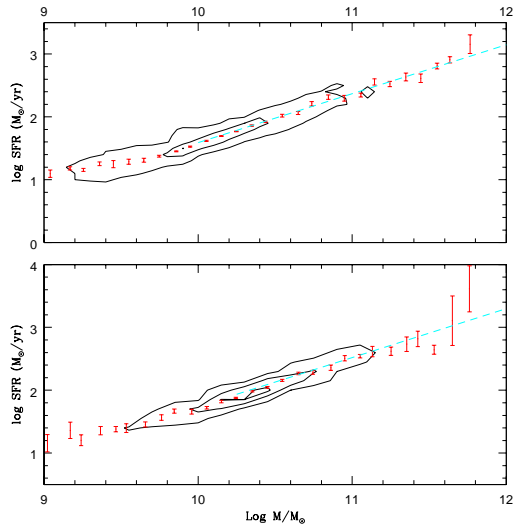}
\caption{The star-forming galaxy main sequence median values (red dots) at (a) $4 < z < 4.8$ and (b) $4.6 < z < 6$, with contours from Fig. 2 superimposed.  There is no decrease in star formation rate or any other evidence of quenching even for the highest-mass star forming galaxies.  That is, the power-law form holds at all masses.} 
\label{fig:SFturnoff}
\end{figure}

At the higher redshifts observed in SPLASH, the SFR main sequence is mass- and SFR-complete for the portion of the distribution indicated in Fig. \ref{fig:SFfig}.   The SFR main sequence is reasonably approximated by a linear relationship for all masses above mass completeness and exhibits no evidence of a high-mass turnoff (Fig. \ref{fig:SFturnoff}).   However, due to the depth of the current optical/NIR data, SPLASH is not sensitive to a population of fully quenched galaxies at this redshift. 

\section{Discussion}
\label{sec:discussion}

Several previous studies have examined the redshift evolution of the SFMS, finding that the slope $\alpha = d(\log SFR)/d(\log M_*)$ may decrease towards higher redshift. SPLASH allows us to check whether this trend continues to $z \sim 6$.  At lower redshift the evolution of $\alpha$ has been fit as a function of redshift \citep{Whitaker2012} and of time \citep{Speagle2014}.  We find the time-based fit is better at high redshift as $(1+z) \propto t^{2/3}$ becomes more clearly non-linear (Table \ref{table:fits}), which makes sense because the physical quantity affecting SFR is likely time, not redshift.  

It is surprising that the SFMS at $4 < z < 6$ has almost exactly the properties that one would extrapolate from lower-redshift behavior, with increasing specific SFRs ($\equiv \psi/M_*$) for $M>10^{11} M_\odot$ galaxies.  This trend cannot continue indefinitely because the virialization time for a galaxy depends upon its initial overdensity, and a typical galaxy with total baryonic mass of $M > 10^{11} M_\odot$ ($M \geq M_*$ ) will not virialize until $z \lesssim 10$, with more massive galaxies having even lower virialization redshifts \citep{Haiman1997}.  Furthermore, there is an effective Eddington limit on star formation for a given galaxy mass before star formation destroys the interstellar medium \citep{Younger2008} of 
\begin{equation}
\psi_{\textrm{max}} = 600 \sigma_{400}^2D_{\textrm{kpc}}\kappa_{100}^{-1}M_\odot\textrm{yr}^{-1},
\end{equation}
where the line-of-sight velocity dispersion in units of 400 km/s ($\sigma_{400}$), the interstellar medium opacity in units of 100 cm$^2$/g ($\kappa_{100}$), and characteristic physical scale of the starburst in kpc ($D_{\textrm{kpc}}$) are likely of order unity.  Thus, even under ideal conditions, it takes at least an additional $\sim 10^8-10^9$ yr beyond virialization for the largest protogalaxies observed in SPLASH to attain $M_* > 10^{11.5} M_\odot$.  The universe at $z\sim 6$ is slightly less than $10^9$ yr old.

At $z \sim 6$ the formation of these galaxies is possible if they spend most of their time at near-maximum SFRs.  Such galaxies would then not have a history of sporadic starbursts, but rather near-continuous high-rate star formation from the moment of collapse until $z=6$. This calculation, however, is only constrained to order of magnitude, so some variability in their SFHs is possible.  However, even under ideal conditions this trend cannot continue far beyond $z > 6$, as there is insufficient time to build up stellar mass.  Thus, we should expect that upcoming Hyper-Suprime-Cam (HSC) data, which is anticipated to extend the SPLASH catalog to $z \sim 8$, should find a turnover redshift where the most massive galaxies are not observed, indicating we are seeing the rapid buildup of galaxies that have just completed their primordial formation and virialization.  Otherwise, a substantially modified theory for early-universe structure formation will be required.

HSC observations will also provide a substantial improvement in completeness in selecting high-mass, low-SFR galaxies. Such galaxies lying only slightly below the SFMS would be present in the current data, but improved Y- and z-band detection using HSC is necessary to select dusty and/or more quiescent high-mass, high-redshift galaxies. The current SPLASH catalog is sufficient to demonstrate that there is no population of turnoff galaxies lying just below the main sequence, but cannot determine whether this is because no galaxies have completed their star formation by $z\sim 6$ or because these galaxies have rapidly changed their SED and are lost due to selection.  Such galaxies have been observed to $z\sim 3.5$ \citep{ilbert+13} and likely form in Eddington limited starbursts \citep{Toft2014}, but it is unclear when the quenching process began. Finding these first turnoff galaxies is essential to understanding when and how the most massive systems in the universe are quenched. 

Taken together, our results provide strong hints that the most massive galaxies at high redshift all have a very similar history, one where they are forming stars at a nearly maximal rate from the first moment of virialization until a rapid turnoff at high redshift.  This might turn out to be incompatible with the current consensus view that most galaxies form their stars through a series of individual starbursts, likely triggered by local events and environment and individual to the history of each galaxy.  However, SPLASH is reporting only on the most massive galaxies, formed at the highest redshifts and very overdense initial fluctuations, so such galaxies are not necessarily typical.  A natural next step is to follow up on this analysis, combining it with observational evidence about the formation process for galactic nuclei to try and build a consistent framework for galaxy evolution (cf.  \citet{Steinhardt2014}), but such a model is beyond the scope of this Letter.

\acknowledgments

The authors would like to thank Steve Bickerton, Sean Carey, Martin Elvis, and Brian Feldstein for helpful comments.

\bibliographystyle{apj}

\end{document}